\documentclass{aa}

\usepackage[varg]{txfonts}
\usepackage{graphicx,amssymb}
\usepackage{epsf,psfig}
\usepackage{natbib}

\begin{document}

\title{Revealing the influence of dark matter on the nature of motion and the families of orbits in axisymmetric galaxy  models}

\titlerunning{Dark matter and galaxy structure}
\authorrunning{E. E. Zotos \& N. D. Caranicolas}

\author{Euaggelos E. Zotos \inst{}
\and
Nicolaos D. Caranicolas \inst{}}

\institute{Department of Physics, Section of Astrophysics, Astronomy \& Mechanics,
Aristotle University of Thessaloniki, \\
GR-541 24, Thessaloniki, Greece \\
\email{evzotos@physics.auth.gr}}

\date{Received 24 July 2013 / Accepted 20 November 2013}

\abstract{An axially symmetric galactic gravitational model composed of a dense, massive and spherical nucleus with an additional dark matter halo component was used, to distinguish between the regular and chaotic character of orbits of stars that move in the meridional plane $(R,z)$. We investigated two different cases: (i) a flat-disk galaxy (ii) an elliptical galaxy. It is of particular interest to reveal how the portion of the dark matter inside the main body of the galaxy influences the ordered or chaotic nature of motion. Varying the ratio of dark matter to stellar mass, we monitored the evolution not only of the percentage of chaotic orbits, but also of the percentages of orbits that compose the main regular resonant families, by classifying regular orbits into different families. Moreover we tried, to reveal how the starting position of the parent periodic orbits of each regular family changes with respect to the fractional portion of dark matter. We compared our results with previous similar work.}


\maketitle

\section{Introduction}
\label{intro}

Over the years, a huge amount of research has been devoted to determining the regular or chaotic nature of orbits in axisymmetric galaxies. However, the vast majority deals mostly with the distinction between regular and chaotic motion, while only a tiny fraction of the existing literature proceeds to classify orbits into different regular families. In a thorough pioneer study \citet{LS92} analyzed the orbital content in the coordinate planes of triaxial potentials and in the meridional plane of axially symmetric potentials, focusing on the regular families. \citet{CA98} developed a method based on the analysis of the Fourier spectrum of the orbit that can distinguish not only between regular and chaotic orbits, but also between loop, box, and other resonant orbits either in two- or three-dimensional potentials. This spectral method was improved and applied in \citet{MCW05} to identify the different types of regular orbits in a self-consistent triaxial model. The same code was improved even more in \citet{ZC13}, who investigated the influence of the central nucleus and of the isolated integrals of motion (angular momentum and energy) on the percentages of orbits in the meridional plane of an axisymmetric galactic model composed of a disk and a spherical nucleus. Recently, \citet{CZ13}, used an analytical dynamical model that describes the motion of stars both in disk and elliptical galaxies that contain dark matter to investigate how the presence and the amount of dark matter influences the regular or chaotic nature of orbits as well as the behavior of the different families of resonant orbits.

\citet{CZ13} represented the main body of the galaxy using the mass model potential proposed by \citet{C12}. We have continue a series of papers begun with \citet{ZC13} and \citet{CZ13} that have as an objective the orbit classification in different galactic gravitational potentials. Here we use a logarithmic potential to model the dynamical properties of the main galaxy. To augment this we compared the two different types of galactic models (mass and logarithmic model) and defined how the particular type of the potential influences not only the amount of chaos, but also the portion of the different families of regular orbits. We clarify that we examined the influence of dark matter enclosed in the main body of the galaxy, but not of the dark-matter halo.

The paper is arranged as follows: In Section \ref{galmod} we present in detail the structure and the properties of our gravitational galactic model. In the following section, we investigate how the parameter corresponding to the fractional portion of the dark matter in galaxies influences the character of the orbits in both disk and elliptical galaxy models. We conclude with Section \ref{discus}, where the discussion and the conclusions of this research are presented.

\section{Presentation and properties of the galactic model}
\label{galmod}

The total potential $V(R,z)$ in our model consists of three components: the main galaxy potential $V_{\rm g}$, the central spherical nucleus $V_{\rm n}$, and the dark matter halo component $V_{\rm h}$. The first one is represented by the new gravitational model proposed by \citet{Z11}. This new model is a combination of the logarithmic and the Miyamoto-Nagai model \citep{MN75} with the addition of the term $(1 + \delta)$. This term $(1 + \delta)$, first used in \citet{C12} and \citet{CZ13}, induces and regulates the amount of dark matter in the main galaxy. The corresponding potential is
\begin{equation}
V_{\rm g}(R,z) = \frac{\upsilon_0^2}{2}\ln \left(R^2 + \left(1 + \delta \right)\left(\alpha + \sqrt{h^2 + z^2}\right)^2 + c^2 \right).
\label{Vg}
\end{equation}
Here, $\delta$ is the fractional portion of the dark matter (or in other words, the ratio of dark matter to luminous/stellar matter) in the main body of the galaxy, while $c$ is a softening parameter. Note that $\delta$ in Eq. (\ref{Vg}) indicates that the dark matter affects the structural parameters of the galaxy. The fractional portion of dark matter has also been inferred by studying the kinematical properties of a set of rotation curves \citep{PS88,PS90,SP99}. The parameter $\upsilon_0$ is a normalization constant and is used for the constancy of the galactic units. The disk or elliptical type of the galaxy is controlled by the parameters $\alpha$ and $h$, which correspond to the horizontal and vertical scale length of the galaxy, respectively. Therefore, this potential allows us to describe a variety of galaxy types from a disk galaxy when $\alpha \gg h$ to an elliptical galaxy when $h \gg \alpha$. Several variants of the logarithmic potential have been used successfully in many previous works to model the properties of an elliptical galaxy in the meridional plane \citep[e.g.,][]{KC00,CP03,Z11}.

To describe of the spherically symmetric nucleus we used a Plummer potential \citep[e.g.,][]{BT08}
\begin{equation}
V_{\rm n}(R,z) = \frac{-G M_{\rm n}}{\sqrt{R^2 + z^2 + c_{\rm n}^2}}.
\label{Vn}
\end{equation}
Here $G$ is the gravitational constant, while $M_{\rm n}$ and $c_{\rm n}$ are the mass and the scale length of the nucleus, respectively. This potential has been used successfully in the past to model and therefore interpret the effects of the central mass component in a galaxy \citep[see, e.g.][]{HN90,HPN93,Z12}. Eq. (\ref{Vn}) is not intended to represent the potential of a black hole, nor that of any other compact object, but just the potential of a dense and massive nucleus therefore, we did not include relativistic effects. The dark-matter halo was modeled by a similar spherically symmetric potential
\begin{equation}
V_{\rm h}(R,z) = \frac{-G M_{\rm h}}{\sqrt{R^2 + z^2 + c_{\rm h}^2}},
\label{Vh}
\end{equation}
where $M_{\rm h}$ and $c_{\rm h}$ are the mass and the scale length of the halo, respectively. The spherical shape of the dark halo is simply an assumption, because galactic halos may have a variety of shapes.

We used a system of galactic units, where the unit of length is 1 kpc, the unit of mass is $2.325 \times 10^7 {\rm M}_\odot$ and the unit of time is $0.9778 \times 10^8$ yr. The velocity unit is 10 km/s, the unit of angular momentum (per unit mass) is 10 km kpc s$^{-1}$, while $G$ is equal to unity. Finally, the energy unit (per unit mass) is 100 km$^2$s$^{-2}$. In these units, the values of the parameters are $c = 2.5$, $M_n = 400$, $c_{\rm n} = 0.25$, $M_{\rm h} = 10000$, and $c_{\rm h} = 25$. For the disk model we chose $\upsilon_0 = 25, \alpha = 3$ and $h = 0.12$, while for the elliptical model we have set $\upsilon = 15$, $\alpha = 0.1$ and $h = 10$. The fractional portion of dark matter, on the other hand, is treated as a parameter and its value varies in the interval $0 \leq \delta \leq 0.5$.

\begin{figure*}
\centering
\resizebox{0.9\hsize}{!}{\includegraphics{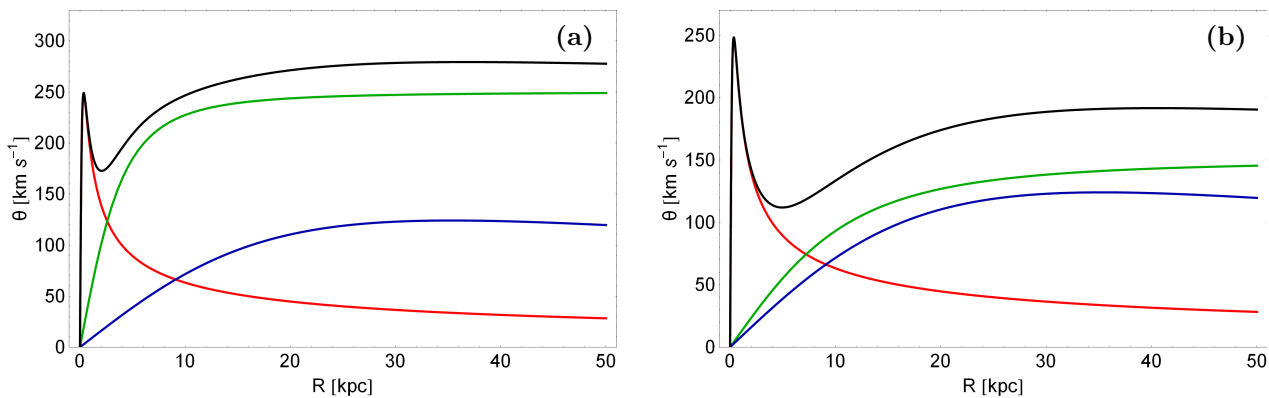}}
\caption{Plot of the rotation curve in our (a-left) disk and (b-right) elliptical galactic model. We can distinguish the total circular velocity (black) and also the contributions from the spherical nucleus (red), the dark matter halo (blue) and that of the galaxy's main body (green).}
\label{rotvel}
\end{figure*}

\begin{figure*}
\centering
\resizebox{0.9\hsize}{!}{\includegraphics{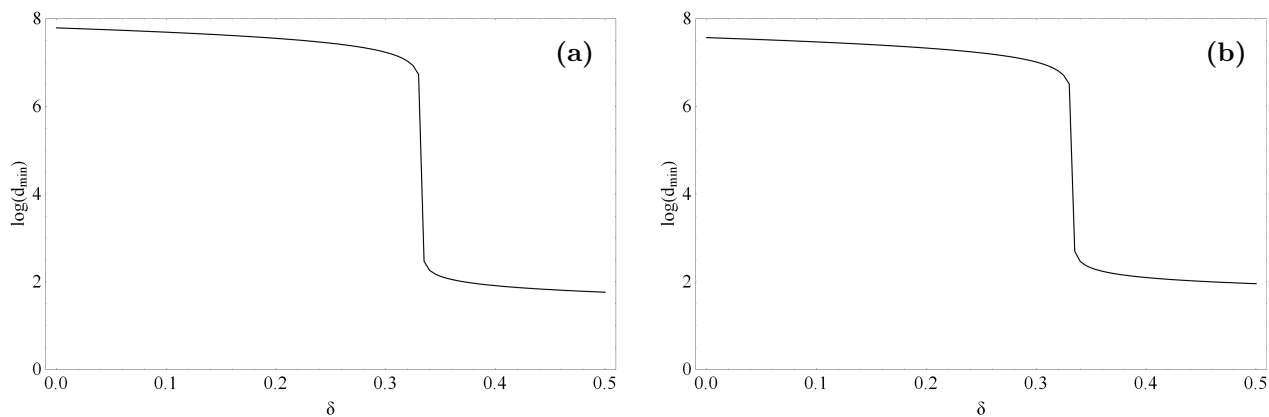}}
\caption{Evolution of the minimum distance $d_{\rm min}$ where negative density appears as a function of the parameter $\delta$ for (a-left) the disk galaxy model and (b-right) the elliptical galaxy model.}
\label{deneg}
\end{figure*}

One very important physical quantity in galaxies is the circular velocity in the galactic plane $(z=0)$, $\theta(R)$. A plot of $\theta(R)$ is presented in Fig. \ref{rotvel}(a-b) as a black curve for the disk and elliptical galactic model when $\delta = 0.5$. Moreover, in the same plot, the red line shows the contribution from the spherical nucleus, the green curve is the contribution from the galaxy main body, while the blue line corresponds to the contribution from the dark matter halo. Each contribution prevails at different distances from the galactic center. In particular, the contribution from the spherical nucleus dominates at small distances, while at large galactocentric distances the contributions from the dark halo and the main galaxy prevail, forcing the rotation curve to remain flat with increasing distance from the center. It is widely accepted that the dark matter contribution to the rotation curve holds high values at large galactocentric distances or even increases with radius, which implies the existence of in-visible matter. It can be seen in Fig. \ref{rotvel}(a-b) that in both disk and elliptical galaxy models the total rotation curve remains flat at large radii. Furthermore, the contribution of the galaxy main body remains flat and higher than that of the dark matter halo (especially in Fig. \ref{rotvel}a). This is because of two reasons: (a) the logarithmic potential used to model the main body of the galaxy has one important characteristic in common with many observed galaxies: it has a flat rotation curve outside its core and at large radii, and (b) according to our hypothesis, the main body of the galaxy (either a disk or elliptical) contains a {\bf{mixture}} of luminous and dark matter (we recall that the total amount of dark matter of the galaxy is inserted not only from the Plummer potential, but also from the logarithmic one). The dark faction in the main body levels out at large radii, so that the dark matter does not dominate in the main body at large radii. For the disk galaxy the circular velocity at large distances is about 280 km/s, while for an elliptical it tends to be 190 km/s. In both cases, the resulting values of the circular velocity are typical values corresponding to disk and elliptical galaxies, respectively. In Fig. \ref{rotvel} we also observe the characteristic local minimum of the rotation curve due to the massive nucleus, which appears at low values of $R$ when fitting observed data to a galactic model \citep[e.g.,][]{GHBL10,IWTS13}.

The mass density in our new galaxy model obtains negative values when the distance from the center of the galaxy described by the model exceeds a minimum distance $d_{min} = \sqrt{R^2 + z^2}$, which strongly depends on the parameter $\delta$. Fig. \ref{deneg}(a-b) shows a plot of $d_{\rm min}$ vs $\delta$ for the both disk and elliptical galaxy models. For $0 \leq \delta < 0.32$ the value of the density is positive everywhere. Only for $\delta > 0.32$ negative values appear and only at extreme values of $z$. In fact, even when $\delta = 0.5$ the areas of negative density occur only when $d_{\rm min} \gtrsim 100$ kpc. However, these distances are far beyond the theoretical boundaries of a real galaxy. Our gravitational potential is truncated at $R_{\rm max} = 50$ kpc for two reasons: (i) if not truncated the total mass of the galaxy modeled by this potential would be infinite, which is obviously not physical, and (ii) this way, we avoid negative density.

Furthermore, taking into account that the total potential $V(R,z)$ is axisymmetric, the $z$-component of the angular momentum $L_z$ is conserved. With this restriction, orbits can be described by means of the effective potential
\begin{equation}
V_{\rm eff}(R,z) = V(R,z) + \frac{L_z^2}{2R^2}.
\label{veff}
\end{equation}
Accordingly, the corresponding Hamiltonian to the effective potential given in Eq. (\ref{veff}) can be written as
\begin{equation}
H = \frac{1}{2} \left(\dot{R}^2 + \dot{z}^2 \right) + V_{\rm eff}(R,z) = E,
\label{ham}
\end{equation}
where $\dot{R}$ and $\dot{z}$ are momenta per unit mass, conjugate to $R$ and $z$, respectively, while $E$ is the numerical value of the Hamiltonian, which is conserved. A more detailed presentation of the equations of motion and the variational equations can be found in the previous papers of this series \citep[e.g.,][]{ZC13,CZ13}.

\begin{figure*}
\centering
\resizebox{0.9\hsize}{!}{\includegraphics{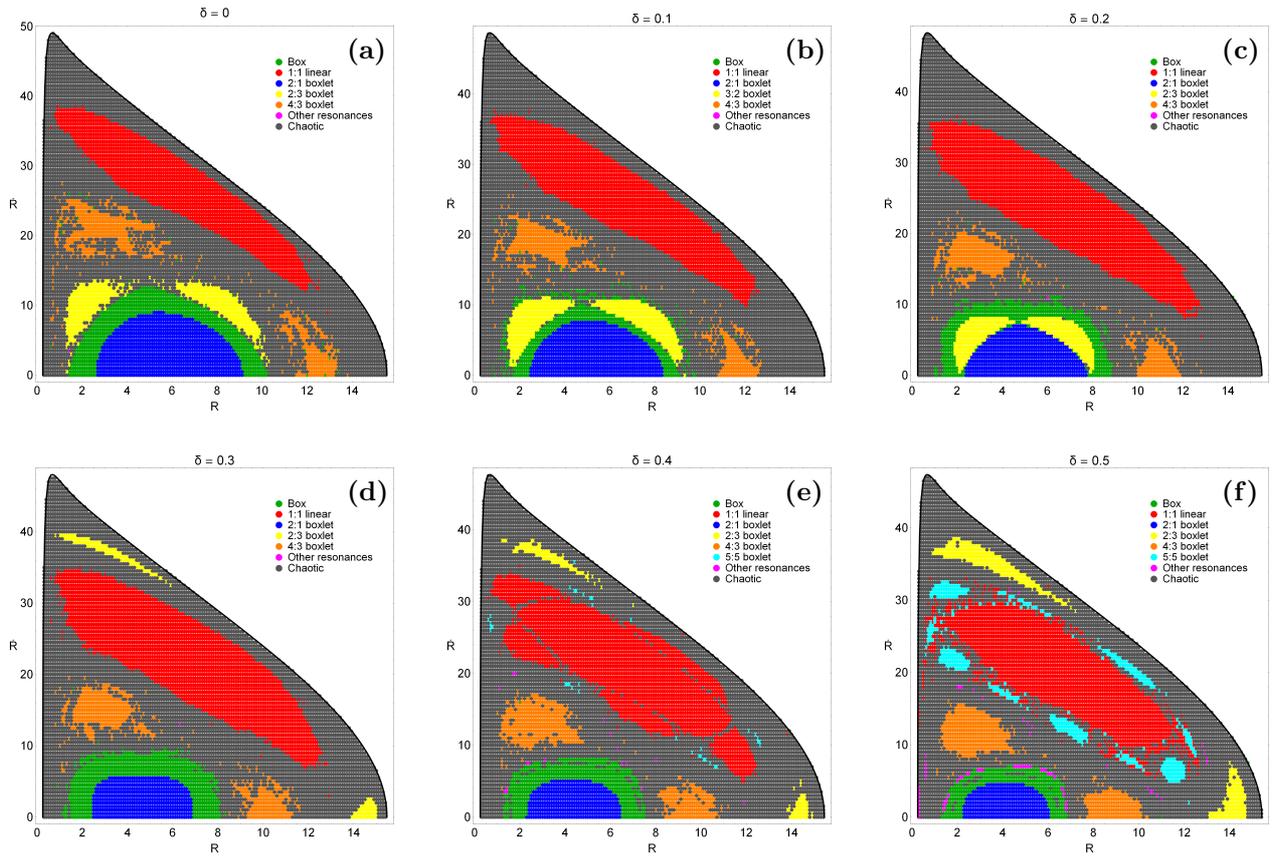}}
\caption{Orbital structure of the $(R,\dot{R})$ phase plane of the disk galaxy model for different values of the fractional portion of the dark matter $\delta$.}
\label{gridsD}
\end{figure*}

\section{Numerical results - orbit classification}
\label{orbclas}

In this section, we numerically integrate several sets of orbits to determine the regular or chaotic nature of motion. For this purpose, we applied the same computational approach as in \citet{CZ13}, that is, the Fast Lyapunov Indicator (FLI) method \citep[e.g.,][]{FGL97,LF01}. In particular, we examined extensive grids of initial conditions $(R_0,\dot{R_0})$ of orbits in every model by numerically integrating the equations of motion and the variational equations. In each case, we computed not only the percentage of chaotic orbits, but also the percentages of orbits composing the main regular resonant families. The classification of regular orbits into different families was made using a method that is based on the field of spectral dynamics. This method calculates the Fourier transform of the coordinates and velocities of an orbit, identifies its peaks, extracts the corresponding frequencies, and then searches for the fundamental frequencies and their possible resonances. In all cases, the value of the angular momentum of the orbits is $L_z = 15$. Moreover, we chose for the disk and elliptical galaxy models energy levels that give $R_{\rm max} \simeq 15$ kpc, where $R_{\rm max}$ is the highest possible value of the $R$ coordinate on the $(R,\dot{R})$ phase plane. After choosing the parameter values, we computed a set of initial conditions and integrated the corresponding orbits by calculating the FLI value and then classifying the regular orbits into different families.

\subsection{Results for the disk galaxy model}

For the disk galaxy models we chose the energy level $E = 1370$, which was kept constant. Our investigation revealed that there are eight main types of orbits in our disk galaxy model: (a) box orbits, (b) 1:1 linear orbits, (c) 2:1 banana-type orbits, (d) 3:2 resonant orbits, (e) 2:3 fish-type orbits, (f) 4:3 resonant orbits, (g) 5:5 resonant orbits, and (h) chaotic orbits. Every resonance $n:m$ is expressed in such a way that $m$ is equal to the total number of islands of invariant curves produced in the $(R,\dot{R})$ phase plane by the corresponding orbit. Most types of orbits are well known and their plots can be found in \citet{ZC13} and \citet{CZ13}.

\begin{figure}
\includegraphics[width=0.9\hsize]{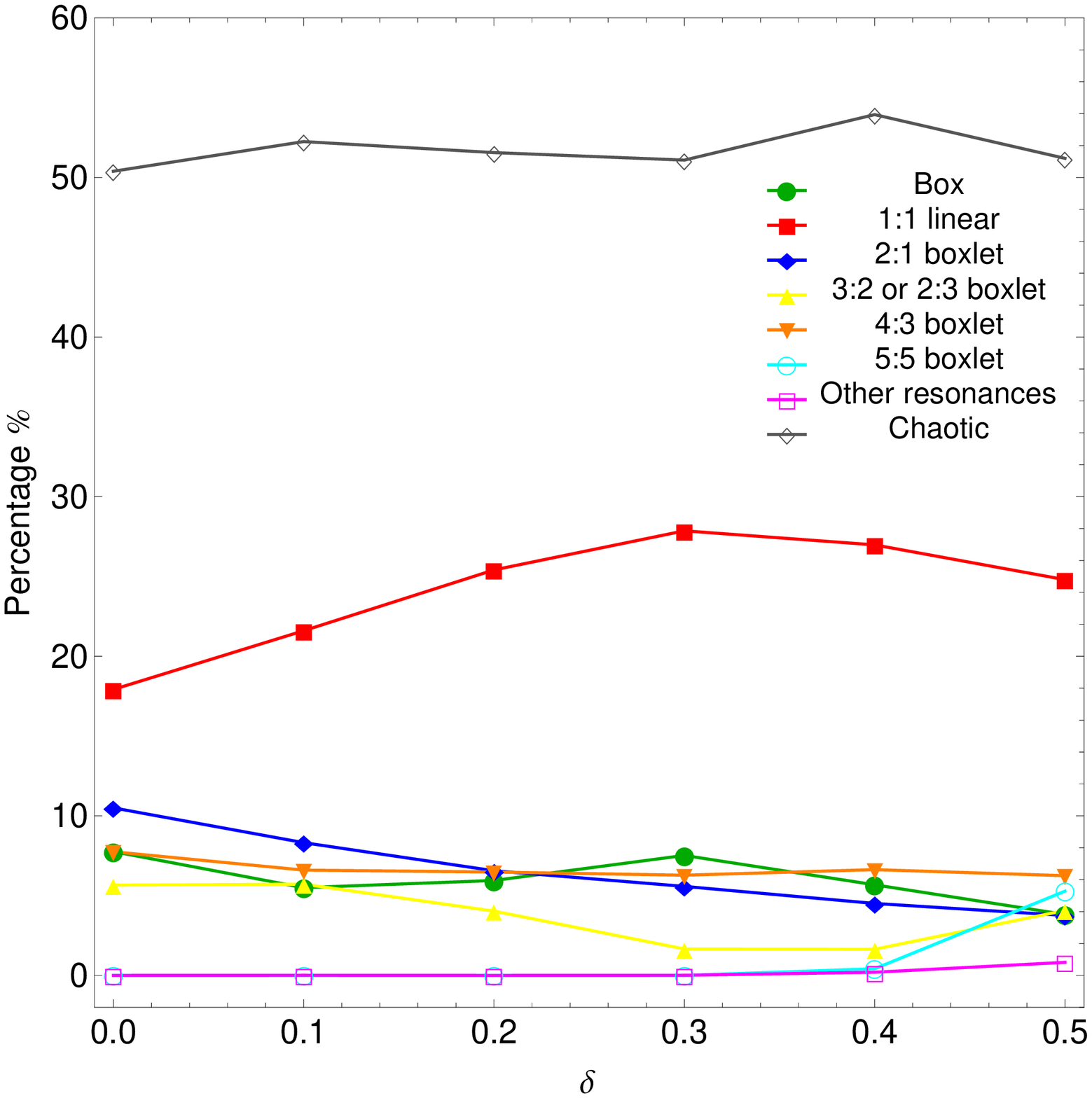}
\caption{Evolution of the percentages of the different types of orbits in our disk galaxy model for varying fractional portions of dark matter $\delta$.}
\label{percsD}
\end{figure}

\begin{figure*}
\centering
\resizebox{0.8\hsize}{!}{\includegraphics{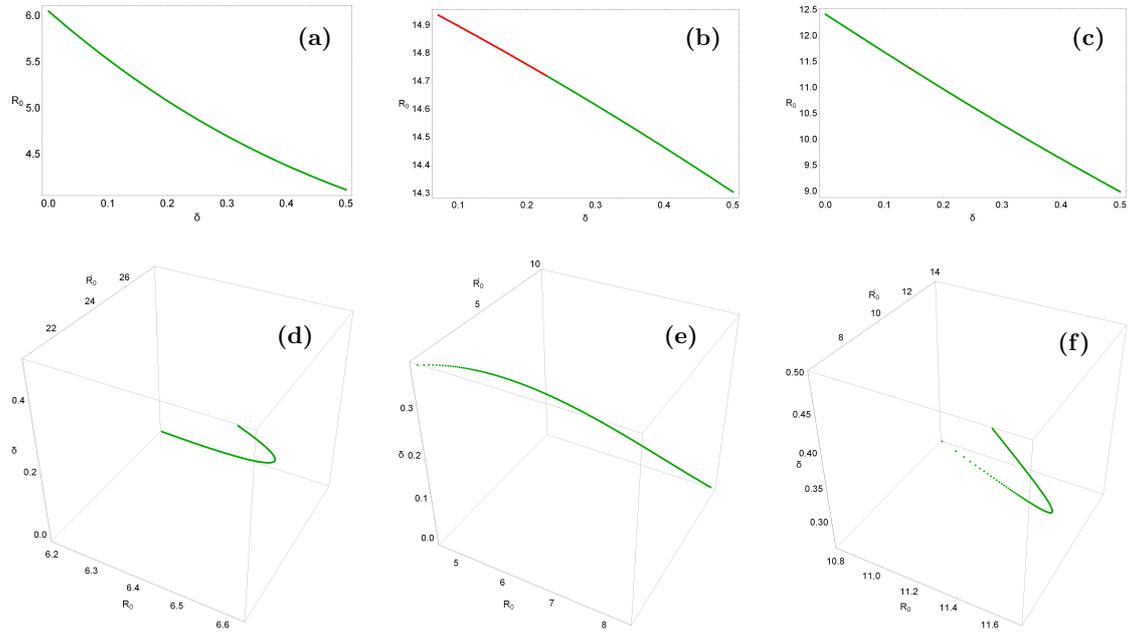}}
\caption{Evolution of the starting position $(R_0,\dot{R_0})$ of the periodic orbits as a function of the fractional portion of dark matter $\delta$. (a) 2:1 resonant family, (b) 2:3 resonant family, (c) 4:3 resonant family, (d) 1:1 resonant family, (e) 3:2 resonant family, and (f) 5:5 resonant family.}
\label{FOPD}
\end{figure*}

To study how the fractional portion of dark matter $\delta$ influences the level of chaos, we let it vary while fixing all the other parameters of our disk galaxy model. We fixed the values of all the other parameters and integrated orbits in the meridional plane for the set $\delta = \{0,0.1,0.2, ..., 0.5\}$. In all cases, the energy was set to $1370$. After choosing the parameter values, we computed a set of initial conditions and integrated the corresponding orbits by computing the FLI of the orbits and then classifying regular orbits into different families.

In Figs. \ref{gridsD}(a-f) we present six grids of orbits that we classified for different values of the fractional portion of dark matter $\delta$. Here, we can identify all the different regular families by the corresponding sets of islands that are formed in the phase plane. In particular, we see the five main families: (i) 2:1 banana-type orbits surrounding the central periodic point, (ii) box orbits are situated mainly outside of the 2:1 resonant orbits, (iii) 1:1 open linear orbits form the double set of elongated islands in the outer parts of the phase plane, (iv) 3:2 resonant orbits form the double set of islands above the box orbits, (v) 2:3 resonant orbits corresponding to the outer triple set of islands shown in the phase plane, (vi) 4:3 resonant orbits forming the inner triple set of islands shown in the phase plane, and (vii) 5:5 resonant orbits producing the set of the five islands around the 1:1 resonance. As expected, apart from the different regions of regular motion, we also observe a unified chaotic sea that surrounds all the islands of stability. The outermost black thick curve is the ZVC. The six grids of initial conditions presented in Fig. \ref{gridsD}(a-f) show that for $\delta > 0.3$ two phenomena occur: (a) the 3:2 resonance disappears, and at the same time the 2:3 resonance appears. Dynamically speaking, a 3:2 orbit is the same as a 2:3 orbit. This is only a matter of which coordinate we put first; were we to shift axes in a $(R,z)$ plot, all $a:b$ orbits would turn into $b:a$, (b) the 5:5 resonant family emerges, which is a bifurcation of the 1:1 family.

The resulting percentages of the chaotic orbits and of the main families of regular orbits as $\delta$ varies are given in Fig. \ref{percsD}. The chaotic motion is always the dominant type of motion, and as the value of $\delta$ varies, the percentage of chaotic orbits exhibits relatively weak fluctuations around 50\%. As the portion of the dark matter increases, there is a gradual increase in the percentage of the 1:1 linear orbits, although their percentage seems to saturate at 25\% when $\delta > 0.3$. The box orbits and the 4:3 resonant orbits, on the other hand, are almost unperturbed by the shifting of the portion of dark matter. Moreover, the meridional 2:1 banana-type orbits exhibit a constant decrease, while the percentage of the miscellaneous orbits (other resonances) remains at very low values. We decided to sum the percentages of 3:2 and 2:3 resonant orbits for the reasons mentioned previously. The percentage of these orbits decreases for $0 \leq \delta < 0.3$, while when $\delta > 0.3$ it remains almost constant at about 5\%. Finally, a substantial increase of the percentage of the bifurcated 5:5 resonant orbits takes place only for $\delta > 0.4$. Therefore, from the diagram shown in Fig. \ref{percsD}, one may conclude that the fractional portion of the dark matter mostly affects the 1:1, 3:2, or 2:3 resonant orbits and chaotic orbits in disk galaxy models.

It is of particular interest to investigate how the variation in the fractional portion of the dark matter influences the position of the different families of periodic orbits shown in the grids of Fig. \ref{gridsD}. For this purpose, we used the theory of periodic orbits described in \citet{MH92} and the numerical algorithm for locating periodic orbits developed and applied in \citet{Z13}. In Fig. \ref{FOPD}(a-f) we present the evolution of the starting position of the parent periodic orbits of the six basic families of resonant orbits. The evolution of the 2:1, 2:3, and 4:3 families shown in Figs. \ref{FOPD}(a-c) is two-dimensional since the starting position $(R_0,0)$ of these families lies on the $R$ axis. In contrast, studying the evolution of the 1:1, 3:2 and 5:5 families of periodic orbits is indeed a real challenge due to the peculiar nature of their starting position $(R_0,\dot{R_0})$. To visualize the evolution of these families, we need three-dimensional plots such as those presented in Figs. \ref{FOPD}(d-f), taking into account the simultaneous relocation of $R_0$ and $\dot{R_0}$.

The stability of the periodic orbits can be obtained using a parameter called the \emph{stability index} \citep{CZ13}. For each set of values of $\delta$, we first located the position of the parent periodic orbits by means of an iterative process. Then, using these initial conditions, we integrated the variational equations to obtain the matrix $X$, with which we computed the index $K$. Our numerical calculations indicate that there are stable and also unstable periodic orbits in the disk galaxy models. In Figs. \ref{FOPD}(a-f) green dots correspond to stable periodic orbits, while red dots correspond to unstable ones. The 2:1, 1:1, and 5:5 periodic orbits remain stable throughout the entire range of the values of $\delta$. On the other hand, for the 2:3 resonance there is a limit of stability at $\delta = 0.2279$. Although there is no evidence of the 5:5 resonance in Fig. \ref{gridsD}d, Fig. \ref{FOPD}f clearly indicates that the resonance is indeed present, although evidently deeply buried in the chaotic sea. The vast majority of the 4:3 periodic orbits are stable, except for those in the region $0.1425 \leq \delta \leq 0.1441$, in which the periodic orbits become unstable. It is clear therefore, that the amount of the dark matter in the galaxy, as well as the isolating integrals of motion, play a fundamental role in the stability of the different regular families, which in turn determines which families are present in each case.

\subsection{Results for the elliptical galaxy model}

For the elliptical galaxy model, we chose the energy level $E = 300$, which was kept constant. Our numerical investigation shows that there are seven main types of orbits in our elliptical galaxy model: (a) box orbits, (b) 1:1 linear orbits, (c) 2:1 banana-type orbits, (d) 3:2 resonant orbits, (e) 4:3 resonant orbits, (f) 8:5 resonant orbits, and (g) chaotic orbits. The basic resonant families, that is, the 2:1, 1:1, 3:2, and 4:3 are common in both disk and elliptical galaxy models. However, for the elliptical galaxy the 8:5 secondary resonant family appears. Again, every resonance $n:m$ is expressed in such a way that $m$ is equal to the total number of islands of invariant curves produced in the $(R,\dot{R})$ phase plane by the corresponding orbit.

\begin{figure*}
\centering
\resizebox{0.9\hsize}{!}{\includegraphics{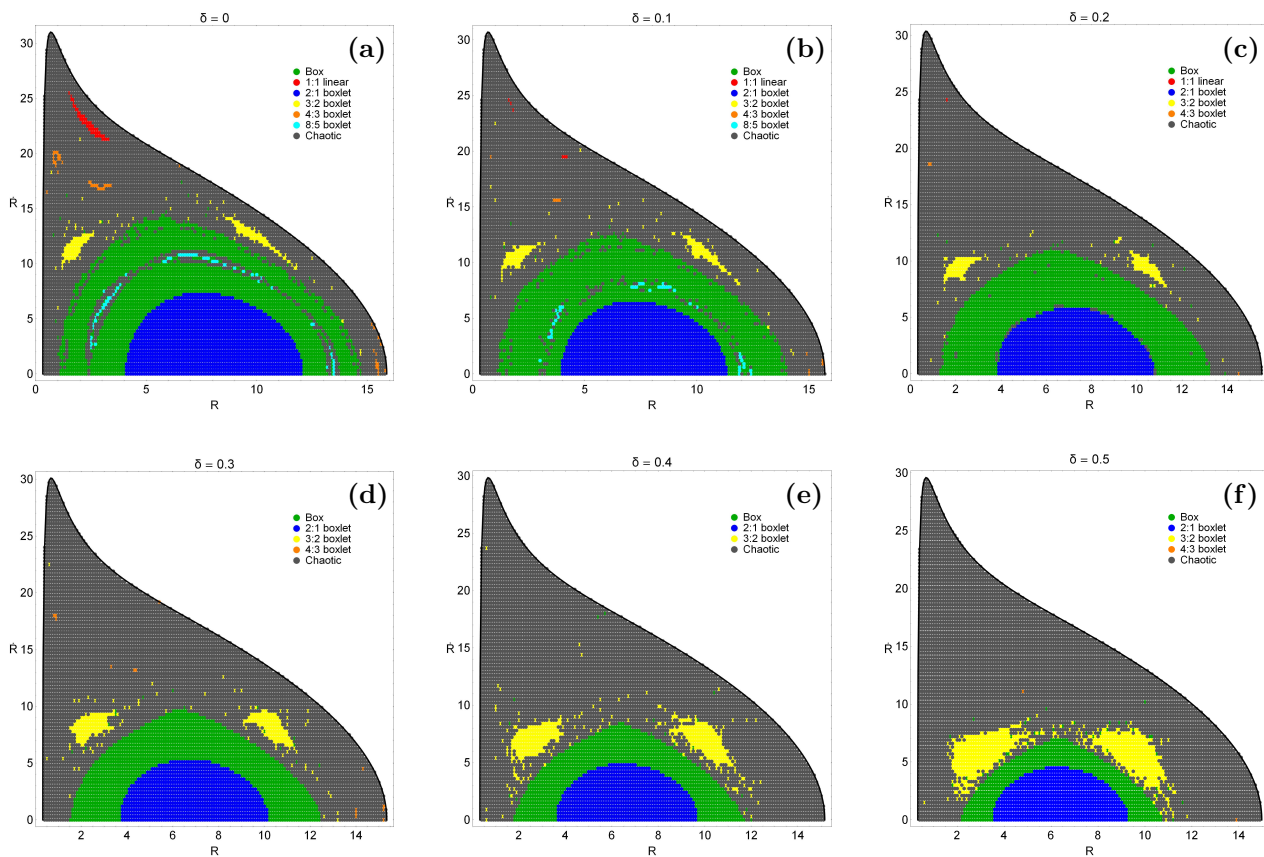}}
\caption{Orbital structure of the $(R,\dot{R})$ phase plane of the elliptical galaxy model for different values of the fractional portion of dark matter $\delta$.}
\label{gridsE}
\end{figure*}

To study how the fractional portion of the dark matter $\delta$ influences the level of chaos, we let it vary while fixing all the other parameters in our elliptical galaxy model. We fixed the values of all the other parameters and integrate orbits in the meridional plane for the set $\delta = \{0,0.1,0.2, ..., 0.5\}$. In all cases the energy was set to $300$.

Six grids of initial conditions $(R_0,\dot{R_0})$ that we classified for different values of the fractional portion of the dark matter $\delta$ are shown in Figs. \ref{gridsE}(a-f). By inspecting these grids, we can easily identify all the regular families by the corresponding sets of islands that are produced in the phase plane. In particular, we see the six main families of orbits already mentioned, which are plotted with different colors: (i) 2:1 banana-type orbits correspond to the central periodic point, (ii) box orbits are situated mainly outside of the 2:1 resonant orbits, (iii) 1:1 open linear orbits form the double set of elongated islands in the outer parts of the phase plane, (iv) 3:2 resonant orbits form the double set of islands, (v) 4:3 resonant orbits correspond to the outer triple set of islands shown in the phase plane, and (vi) 8:5 resonant orbits producing the set of the five small islands. This clear shows that the structure of the phase plane in the elliptical galaxy models differs greatly from that of the disk models. In all cases an immense chaotic sea dominates the phase plane, indicating that the vast majority of orbits move in chaotic orbits. Moreover, there is an inner weak chaotic layer inside the area of the box orbits, for $\delta < 0.2$. This indicates that there is no unified chaotic domain, at least at the $z = 0$ slice of the phase space. The 8:5 resonance is located at this weak chaotic layer. In fact, in galaxy models with enough dark matter ($\delta \geq 0.2$) this chaotic layer disappears from the phase plane which results to the extinction of the 8:5 resonant orbits. As the value of $\delta$ increases, essential types of orbits (i.e., the 1:1 linear orbits) are rapidly depleted, giving place to chaotic orbits. Furthermore, the islands of the 4:3 resonant orbits are so small and deeply buried in the thick chaotic sea that they appear as lonely points in the grid. With a much closer look, we see that the allowed radial velocity $\dot{R}$ of the stars moving near the center of the galaxy is about 300 km/s which is considerably lower than for the disk galaxy models (about 500 km/s).

\begin{figure}
\includegraphics[width=0.9\hsize]{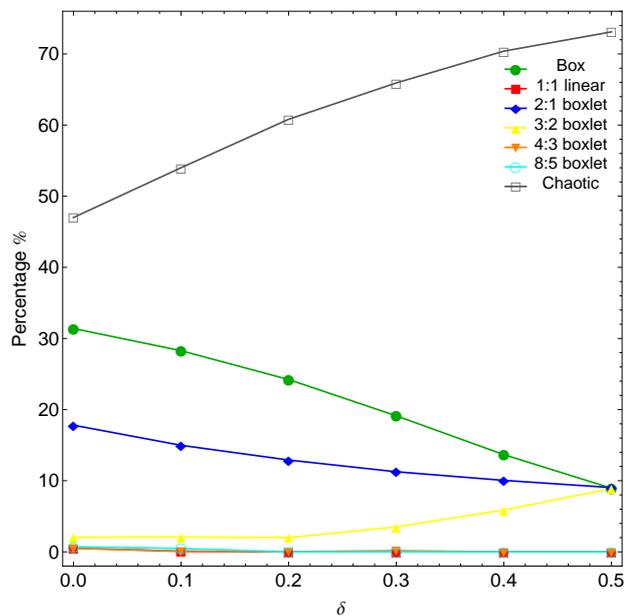}
\caption{Evolution of the percentages of the different types of orbits in our elliptical galaxy model for varying fractional portions of dark matter $\delta$.}
\label{percsE}
\end{figure}

In Fig. \ref{percsE} we present the resulting percentages of the chaotic orbits and of the main families of regular orbits as $\delta$ varies. The motion of stars in elliptical galaxies is highly chaotic throughout. In particular, as the portion of dark matter is increased, the percentage of chaotic orbits exhibits an almost linear growth. In contrast, at the same time the percentages of the meridional 2:1 orbits and the box orbits start to drop along a linear reduction. On the other hand, the 3:2 family is the only family that slowly but constantly amplifies its percentage with increasing $\delta$. The remaining families of orbits change very little; the 4:3 and the 8:5 orbits are almost unperturbed by the shifting of the amount of dark matter and present a monotone evolution with extremely low percentages $(< 1\%)$. At the higher value of the portion of dark matter $(\delta = 0.5)$, the percentages of the box, 2:1 and 3:2 orbits tend to a common value (around 10\%), thus sharing three tens of the entire area of the phase plane. From Fig. \ref{percsE} one may conclude that dark matter in elliptical galaxies mostly affects box, 2:1, 3:2, and chaotic orbits. In fact, a large portion of box and 2:1 orbits turns into chaotic as the galaxy gains more dark matter.

\begin{figure*}
\centering
\resizebox{0.8\hsize}{!}{\includegraphics{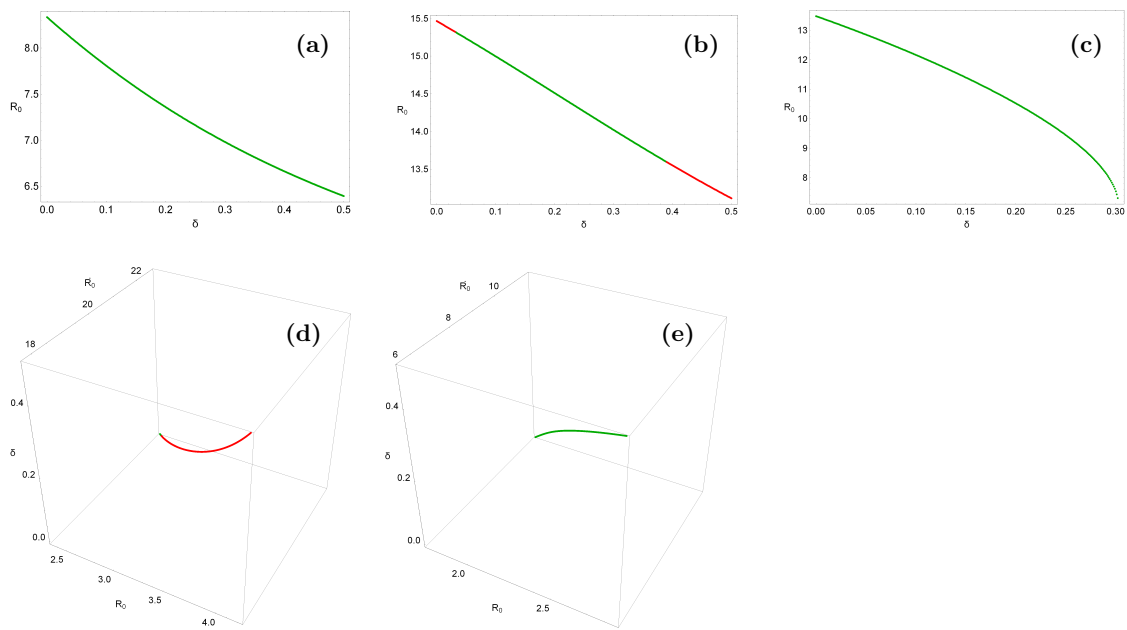}}
\caption{Evolution of the starting position $(R_0,\dot{R_0})$ of the periodic orbits as a function of the fractional portion of dark matter $\delta$. (a) 2:1 resonant family, (b) 4:3 resonant family, (c) 8:5 resonant family, (d) 1:1 resonant family, and (e) 3:2 resonant family.}
\label{FOPE}
\end{figure*}

We continue with the influence of the variation in the fractional portion of dark matter on the position of the different families of periodic orbits shown in the grids of Fig. \ref{gridsE}(a-f). To facilitate this, we followed the same method as in the case of the disk galaxy. In Fig. \ref{FOPE}(a-e) the evolution of the starting position $(R_0,\dot{R_0})$ of the parent periodic orbits of the five basic families of resonant orbits is given. Again, the evolution of the 2:1, 4:3, and 8:5 families shown in Figs. \ref{FOPE}(a-c) is two-dimensional since the starting position $(R_0,0)$ of these families lies on the $R$ axis. On the other hand, the evolution of the 1:1 and 3:2 families are shown in the three-dimensional plots in Figs. \ref{FOPE}(d-e) which simultaneously follow the relocation of $R_0$ and $\dot{R_0}$ as the fractional portion of dark matter $\delta$ varies. In Figs. \ref{FOPE}(a-e) green dots correspond to stable periodic orbits, while red dots correspond to unstable ones. The 2:1, 3:2, and 8:5 periodic orbits remain stable throughout the entire range of the values of $\delta$. On the other hand, for the 4:3 resonance there is an interplay between stable and unstable periodic orbits. To be precise, for $0 \leq \delta \leq 0.0310$, and $0.3910 \leq \delta \leq 0.5$ the 4:3 periodic orbits are unstable, while they are stable for all the intermediate values of $\delta$. The vast majority of the 1:1 resonant periodic orbits are unstable and stable 1:1 periodic orbits are present only at the interval $0 \leq \delta \leq 0.0120$. This explains why the 1:1 family is absent from the grids of Figs. \ref{gridsE}(d-f). Even though the 8:5 family exists until $\delta = 0.3022$, there is no evidence of the 8:5 resonance in the corresponding grids of Figs. \ref{gridsE}(c-d). A logical assumption justifying this phenomenon would be that the 8:5 resonance corresponds to extremely small islands of stability in the phase plane, that remain undetected in our grids, and only if we increase the density of the grid significantly will they appear. Therefore it is clear that the amount of the dark matter in elliptical galaxy models influences not only the existence, but also the stability of the different regular families.

\section{Discussion and conclusions}
\label{discus}

We used an analytic, axisymmetric galactic gravitational model that embraces the general features of a disk or elliptical galaxy with a dense, massive nucleus and a spherical dark matter halo component. To facilitate our study, we chose to work in the meridional plane $(R,z)$. Our aim was to investigate how influential the parameter is that corresponds to the portion of dark matter $\delta$ on the level of chaos and also on the distribution of regular families in disk and elliptical galaxy models. Our results suggest that the level of chaos and the distribution in regular families is indeed very dependent on the portion of dark matter in galaxies. We kept the values of all the other parameters constant because our main objective was to investigate the influence of the portion of dark matter $\delta$ on the percentages of the orbits. In a previous paper \citep{ZC13} we showed that the isolating integrals of motion (angular momentum and energy) significantly affect the orbital structure. Therefore, we felt that we should not re-investigate the influence of the isolating integrals in this case.

We observed that in disk galaxy models, chaotic motion is always the prevailing type of motion, however, the chaotic percentage is little affected by the increase of the portion of dark matter. In fact, chaotic orbits cover about half of the phase plane throughout the range of the values of $\delta$. In models with enough of dark matter ($\delta > 0.3$), interesting phenomena regarding the orbital structure take place. The 3:2 resonance disappears and is replaced by the 2:3 resonance; its dynamical sibling. Moreover, additional bifurcated resonant families (i.e., the 5:5 family, which is a bifurcation of the basic 1:1 family) start to emerge, exhibiting substantial increase in their rates for $\delta > 0.4$. In the interval $0 \leq \delta < 0.3$ we observed a gradual increase in the percentage of the 1:1 linear orbits at the expense of the 2:1 banana-type and 3:2 orbits. On the other hand, all the remaining regular families are almost unperturbed by the shifting of the portion of dark matter.

We found that in elliptical galaxy models the motion of stars is highly chaotic. In particular, the percentage of chaotic orbits grows sharply as dark matter is being accumulated, while at the same time the 2:1 meridional banana-type and box orbits are depopulated following almost a linear decrease at their percentages. The only family that amplifies its rate with increasing amount of dark matter is the 3:2 family, while on the other hand, all the remaining families of regular orbits change very little, having extremely low percentages. We also observed that at the higher value of the portion of dark matter, that is for $\delta = 0.5$, the percentages of the box, 2:1, and 3:2 orbits tend to reach a common value around 10\%. In elliptical galaxy models with a low portion of dark matter $(\delta < 0.2)$ a weak chaotic layer exists inside the region of box orbits. The secondary resonant orbits 8:5 live and grow inside this chaotic layer. Nevertheless, the main chaotic sea is so thick and extended that the islands of stability in the phase plane that correspond to basic families such as the 1:1 and 4:3 are hardly visible.

The particular type of the potential we used to model the properties of the main galaxy model was a choice of paramount importance. This became more than evident, when we compared the current results, where a logarithmic potential was used for the main galaxy, with the corresponding outcomes derived in a previous work, \citet{CZ13} where the main galaxy body was modeled using a mass-model potential. The main points indicating the similarities and differences between the two types of potentials can be summarized as follows:

\begin{enumerate}
  \item In disk galaxy models, where the main galaxy is modeled by a logarithmic potential, the fractional portion of the dark matter mostly affects the 1:1, 3:2, or 2:3 resonant orbits and chaotic orbits, while when a mass model is used, the 1:1, 4:3 resonant orbits and the chaotic orbits were more influenced.
  \item The portion of dark matter in elliptical galaxy models mainly influences the box, 2:1, 3:2, and chaotic orbits when the main galaxy is described by a logarithmic potential, but when a mass model is used for the same purpose, only box and chaotic orbits are affected by dark matter, while all the resonant families are almost unaffected by dark matter.
  \item When using a mass model for the main galaxy we observed, in elliptical galaxy models with a low portion of dark matter the motion was entirely regular, and box orbits were the all-dominant type. In contrast, the motion of stars in elliptical galaxies was found to be highly chaotic when a logarithmic potential was used.
  \item The type of the potential also affects the stability of the families of regular orbits. We saw in Section \ref{orbclas} that the portion of dark matter substantially affects the stability of the different families of orbits, hinting at a deep interplay between chaos and proportion of regular families. On the other hand, when the main galaxy was modeled using a mass model, this influence disappeared completely and all periodic orbits remained stable throughout the range of values of $\delta$.
  \item In both types of potentials, negative values of density exist mainly when the portion of dark matter is high (reaching the limit value of 0.5). In the case of the mass model, negative density started to appears at galactocentric distances on the order of about 40 kpc. In contrast, when we used a logarithmic potential to model the main body galaxy, negative density occurred at extremely large distances, at about 100 kpc therefore, while the limit of a real galaxy is about 50 kpc, we may argue that when one uses a logarithmic potential, negative density is practically absent.
\end{enumerate}

We consider the results of the present research as an initial effort and also a promising step in the task of exploring the orbital structure of disk and elliptical galaxies that contain dark matter. Taking into account that our outcomes are encouraging, we plan to properly modify our dynamical model to expand our investigation in three dimensions, thus obtaining the entire network of periodic orbits, revealing the evolution of their stability, and computing the evolution of the percentages of the regular families with respect to the amount of dark matter.

\end{document}